\RequirePackage{amsmath}

\documentclass[preprint,3p,number,sort&compress,11pt,twocolumn]{elsarticle}
\usepackage{color}
\usepackage{amsmath}
\usepackage{amssymb}
\usepackage[hidelinks]{hyperref}
\usepackage[]{units}
\usepackage{graphicx}
\usepackage{caption}
\usepackage{tabularx}
\renewcommand{\vec}[1]{\mathbf{#1}}

\usepackage{setspace}

\usepackage[english]{babel}
\usepackage[T1]{fontenc}
\usepackage[utf8]{inputenc}
\usepackage{lmodern}

\newcommand{\MU}[1]{#1}

\newcommand{\ctoa}{\left(\frac{c}{a}\right)}
\newcommand{\cta}[1]{\left.\frac{c}{a}\right\rvert_{\mathrm{NM}}^{\scriptscriptstyle \mathrm{#1}}}

\newcommand{\ctam}[1]{\left.\frac{c}{a}\right\rvert_{\mathrm{M}}^{\scriptscriptstyle \mathrm{
#1}}}

\newcommand{\atc}[1]{\left.{\frac{a}{c}}\right\rvert_{\mathrm{\scriptscriptstyle NM}}^{
\scriptscriptstyle #1}}

\newcommand{\ax}[2]{{a}_{\mathrm{\scriptscriptstyle #1}}^{
\mathrm{\scriptscriptstyle #2}}}

\newcommand{\bx}[2]{{b}_{\mathrm{\scriptscriptstyle #1}}^{
\mathrm{\scriptscriptstyle #2}}}

\newcommand{\cx}[2]{{c}_{\mathrm{\scriptscriptstyle #1}}^{
\mathrm{\scriptscriptstyle #2}}}

\newcommand{\gx}[2]{\gamma_{\mathrm{\scriptscriptstyle #1}}^{\mathrm{
\scriptscriptstyle #2}}}

\newcommand{\vx}[3]{\vec{#1}_{\mathrm{\scriptscriptstyle #2}}^{
\mathrm{\scriptscriptstyle #3}}}

\addto\captionsenglish{
  \renewcommand{\contentsname}%
    {}%
}

\begin{document}

\begin{frontmatter}

\title{Geometry of Adaptive Martensite \MU{in Ni-Mn-based Heusler alloys}}

\author[1]{Robert Niemann}

\author[1,2]{Sebastian Fähler\corref{mycorrespondingauthor}}
\cortext[mycorrespondingauthor]{Corresponding author}
\ead{s.faehler@ifw-dresden.de}

\address[1]{IFW Dresden, P.O. Box 27 01 16 D-01171 Dresden, Germany}
\address[2]{Technische Universität Dresden, Institut für Festkörperphysik, Dresden, Germany}

\begin{abstract}
Modulated martensites play an important role in magnetic shape memory alloys, because all functional properties are closely connected to the twin microstructure and the phase boundary. The nature of the modulated martensites is still unclear. One approach is the concept of adaptive martensite, which regards all modulated phases as nanotwinned microstructures. In this article, we \MU{use the Ni-Mn-based shape memory alloys as an example to} show the geometric rationale behind this concept using analytic equations based on the phenomenological theory of martensite. 
This could enhance discussions about the implications of the adaptive martensite by showing the exact relations between the various unit cells used to describe the structure. We use the concept to discuss the compatibility at the habit plane, the nature of high-order twin boundaries and the dependence of the lattice constants on the different types of modulation.
\end{abstract}

\end{frontmatter}



\section{Introduction}
In magnetic shape memory alloys, the functional properties are closely connected to the structure and microstructure \cite{Bhattacharya2005}. Especially the modulated phases of martensite play a crucial role in these types of alloys \cite{Niemann2012a}. The prototypical $\mathrm{Ni_2MnGa}$ is well known for giant magnetic-field-induced strains mediated by the movement of twin boundaries \cite{Ullakko_APL96,Sozinov2013}. The stress or magnetic field necessary to move twin boundaries depends strongly on their type \cite{Straka2011b,Heczko2013,Heczko2013a,Seiner2014}. All isostructural Heusler alloys based on Ni-Mn-X (X=Ga, In, Sn, Al, Sb) also show magnetocaloric \cite{Krenke2005} and elastocaloric \cite{Bonnot2008} effects. The reversibility of these effects is hampered by a large hysteresis, which is a consequence of  nucleation and phase boundary movement. Understanding the microstructure at the phase boundary is an important step in order to reduce hysteresis and design materials for magnetocaloric refrigeration \cite{Faehler2012}. A narrow hysteresis also leads to a reduced functional fatigue \cite{Chluba2015}.
\begin{figure}[tbp]
	\centering
		\includegraphics[width=\columnwidth]{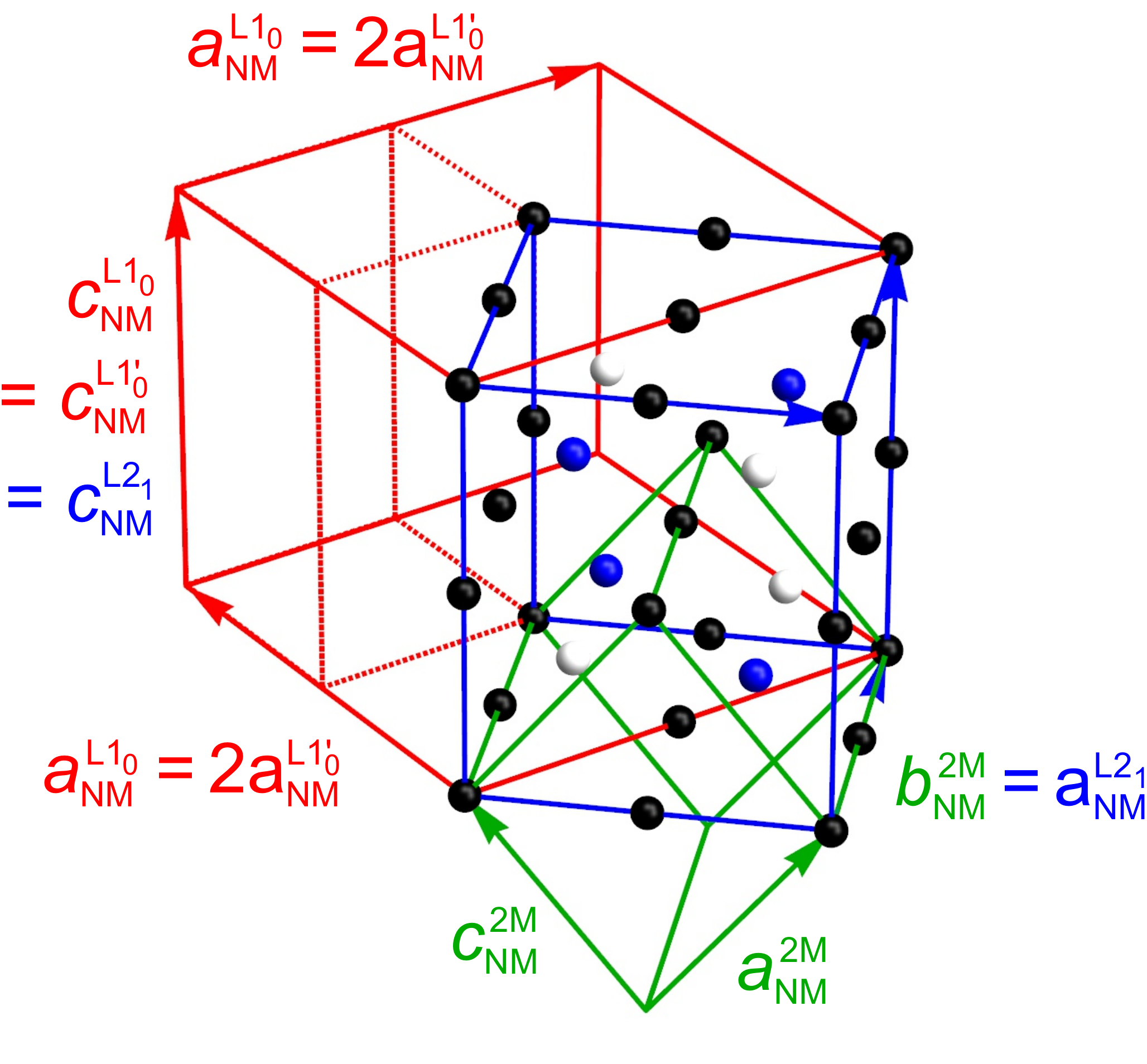}
	\caption{The NM in the $\mathrm{L}2_1$-description ($\cta{\mathrm{L}2_1}>1$) derived from the cubic lattice (blue). In the $\mathrm{L}1_0$ description, there is a larger tetragonal cell with $1>\cta{L1_0}>1/\sqrt{2}$ (red) and a smaller (red, dashed) with $\cta{L1'_0}>\sqrt{2}$. In green, the monoclinic "2M"-description is shown. }
	\label{fig:Fig1}
\end{figure}\MU{
To describe the nature of the modulated phases, the concept of adaptive martensite was first proposed by Khachaturyan et al. \cite{Khach1991a} and then applied to the Ni-Mn-Ga system \cite{Kaufmann_PRL10, Kaufmann2011}. In this concept, the modulated structures are described as nanotwinned microstructures of the tetragonal non-modulated (NM) phase that form to minimize the elastic energy at the phase boundary. Alternative models base on Fermi surface nesting \cite{Johannes2008, Benesova2015} and phonon softening \cite{Schubert2015} as driving mechanism. A unified description of both concepts, which is based on the ordering of nanotwin boundaries, was proposed recently \cite{Gruner2016}.} 
Acceptance of the concept of adaptive martensite was limited in the community also because the geometry becomes very complex and that a confusing variety of different nomenclatures and unit cells are used in the literature. The previous descriptions of adaptive martensite used simplified geometrical models. This article aims at an exact, analytical description of the geometry, which should enhance future critical discussions about the concept of adaptive martensite in the community. 

In Ni-Mn-X, mostly \MU{the modulated phases} 14M, 10M \cite{Pons2000}, and less often 4O are reported \cite{Lin2016}, though it is energetically favorable \cite{Zeleny2016}. Often, a mixture between the modulated phases and NM is observed \cite{Erkartal2012}. 

Because of the large variety of descriptions for the martensitic phases in Ni-Mn-X and the complex geometry of the construction of the adaptive phase, this work aims to present analytical calculations and realistic pictures of the actual geometry of the modulated phases and of the phase boundary. This could serve as a basis for a better understanding between researchers with different backgrounds working on the same material class. 

Special attention is also given to $a$-$b$-twins, which seem to play an important role in twin boundary mobility \cite{Heczko2013}. They can be constructed using the concept of nanotwinning, because \MU{this compound} twin boundary can be parallel to the nanotwin boundaries. 

The formation of nanotwins is driven by the elastic stress at the phase boundaries in combination with an extremely low twin boundary energy \cite{Khach1991a,Kaufmann_PRL10}. This energy balance favors the introduction of twins in the non-modulated martensite on the length scale of the lattice parameter. Consequently, the concept of adaptive martensite via nanotwinning uses the tetragonal, non-modulated martensite as the fundamental structure to build a modulated lattice. If volume conservation is assumed, the $\cta{}$-ratio is the main parameter necessary to describe the entire structure and also the microstructure of the modulated phases, which usually has several orders of hierarchy \cite{Roytburd_93,Kaufmann2011}.  

This article is structured as follows: First, the geometry of the nonmodulated and the modulated lattice is discussed with respect to the different descriptions of the martensitic unit cells found in the literature. Analytical expressions are achieved to find all atomic positions and characteristic angles. These expressions are used to plot the dependence of the lattice parameters as a function of the tetragonality of NM.  
The same expressions are employed to sketch the 10M, 14M and 4O lattices in real space as a function of $\ctoa$. High resolution GIF-animations can be found in the supplemental material \cite{Supp1}. 
Finally, the orientation of the interface between austenite and martensite is calculated and represented graphically for 10M, 14M and 4O in order to give a realistic representation of the misfit at the habit plane. 


\section{The austenite lattice}
The coordinate system used here is
\begin{equation}
\vec{e}_1=\begin{pmatrix}1\\0\\0\end{pmatrix}\quad\vec{e}_2=\begin{pmatrix}0\\1\\0\end{pmatrix}\quad\vec{e}_3=\begin{pmatrix}0\\0\\1\end{pmatrix}.
\end{equation}
The austenite lattice is given by
\begin{equation}
\vec{a}_1=a_0\cdot \vec{e}_1\quad \vec{a}_2=a_0\cdot \vec{e}_2\quad \vec{a}_3=a_0\cdot \vec{e}_3
\end{equation}
in the $\mathrm{L}2_1$ description, also called the cubic lattice. The full unit cell and chemical ordering are shown e.g. by Webster \cite{Webster1969}. The volume of the austenite unit cell is ${a_0}^3$. 

\section{Variants of NM martensite}
Here, the unit cells of NM are described. Their twin relation forms the basis of the following description of modulated martensite. 

\subsection{The non-modulated lattice}
For non-modulated martensite, different unit cells are used (Fig. \ref{fig:Fig1}). Often, it is described using the $\mathrm{L}2_1$ notation obtained by a linear deformation of the cubic lattice. The $\cta{\mathrm{L}2_1}$ is larger than 1. Typical lattice constants are $a_\mathrm{\scriptscriptstyle NM}^{\scriptscriptstyle \mathrm{L}2_1}=0.54\,\mathrm{nm}$, $c_\mathrm{\scriptscriptstyle NM}^{\scriptscriptstyle \mathrm{L}2_1}=0.66\,\mathrm{nm}$.
Also the $\mathrm{L}1_0$ description is used, e.g. by Pons et al \cite{Pons2000}. The $\cta{L1_0}$-ratio is smaller than one, but larger than $1/\sqrt{2}$. 
A smaller version of this $\mathrm{L}1_0$ cell is very common and effective e.g. for DFT calculations. \MU{This cell, called $\mathrm{L}1'_0$ here,} is only a quarter of the large $\mathrm{L}1_0$ cell described above. 

Finally, the NM can be described by a monoclinic unit cell \cite{Pond2012}, called "2M". This cell is useful to construct the stacked sequence of modulated martensite. The relations between the lattice constants of all four descriptions are given in Tab. \ref{tab:L10}.

\begin{table*}[t]
\centering
\caption{Alternative descriptions of NM and their relation to the NM-$\mathrm{L}2_1$ parameters. Additionally, the symmetry (Sym.) and the range of the $\ctoa$-ratio in the respective system is given.}
\label{tab:L10}
\begin{tabular}{cccc}
 &  $\mathrm{L}1_0$ & $\mathrm{L}1'_0$ & $2\mathrm{M}$ \\ \hline
Sym. & tetragonal
 & tetragonal & monoclinic \\ 
$a$ &  $\ax{NM}{\mathrm{L}1_0}= \sqrt{2} \ax{NM}{\mathrm{L}2_1}$ & $\ax{NM}{L1'_0}=\frac{1}{\sqrt{2}} \ax{NM}{\mathrm{L}2_1}$ &$\ax{NM}{2M} = \frac12 \sqrt{{\ax{NM}{\mathrm{L}2_1}}^2+{\cx{NM}{\mathrm{L}2_1}}^2}$ \\
$b$ & - & - & $\bx{NM}{2M} = \ax{NM}{\mathrm{L}2_1}$  \\
$c$ & $\cx{NM}{L1_0}=\cx{NM}{\mathrm{L}2_1}$ & $\cx{NM}{L1'_0}=\cx{NM}{\mathrm{L}2_1}$ & $\cx{NM}{2M}=\ax{NM}{2M}$ \\
$\gamma$ &  - & - & $\gx{NM}{2M}=2\tan^{-1}{\left(\atc{\mathrm{L}2_1}\right)}$\\
$\ctoa$ &$\frac{1}{\sqrt{2}}<\cta{L1_0}<1$ & $\cta{L1'_0}\geq\sqrt{2}$ & $\cta{2M}=1$  
\end{tabular}
\end{table*}

\subsection{Lattice stretch and twin formation}
 In the following, only the $\mathrm{L}2_1$ description is used for NM if not stated otherwise, therefore
\begin{equation}
a\equiv \ax{NM}{L21}\qquad c\equiv \cx{NM}{L21}
\end{equation}
The tetragonality of the NM unit cell is given by $\ctoa\geq1$. The tetragonality $\ctoa$ is the main structural parameter, also for the modulated phases. Assuming volume conservation during the austenite to NM transformation, the lattice constants $a$, $c$ of NM are described as functions of $a_0$ and $\ctoa$.

\begin{equation}
a=\sqrt[\leftroot{-1}\uproot{2}\scriptstyle 3]{\frac{V}{\ctoa}}\qquad c={\ctoa} a
\end{equation}

\begin{figure}[tbp]
	\centering
		\includegraphics[width=\columnwidth]{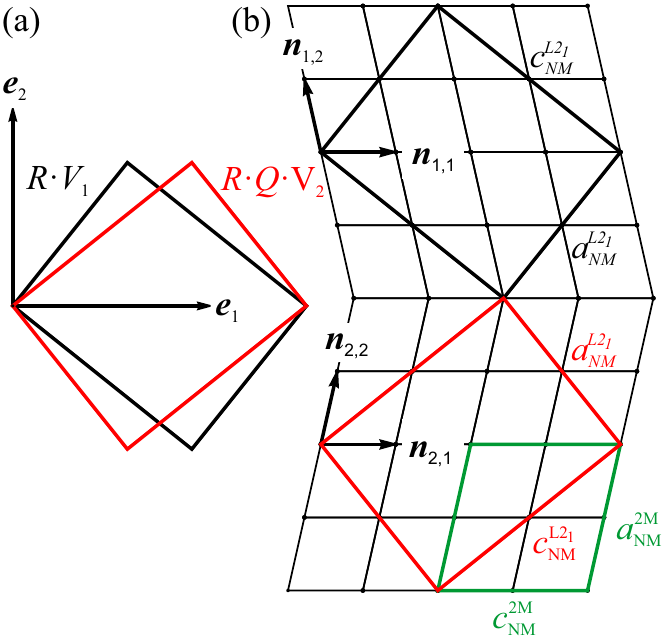}
	
\caption{\label{fig:Fig2}(a) Twin-related orientation of two NM variants. $Q$ rotates variant 2 relative to variant 1; $R$ rotates the entire lattice in order to make the twin boundary parallel to $\vec{e}_1$. (b) Comparison between the twin-related NM unit cells in $\mathrm{L}2_1$-notation in a lattice of NM according to a 2M-based description.}
\end{figure}

The variants of NM can be expressed by the strain matrices

\begin{align}
&V_1=\begin{pmatrix}
\frac{c}{a_0} & 0 & 0\\
0& \frac{a}{a_0}&0\\
0&0&\frac{a}{a_0}
\end{pmatrix}\qquad V_2=\begin{pmatrix}
\frac{a}{a_0} & 0 & 0\\
0& \frac{c}{a_0}&0\\
0&0&\frac{a}{a_0}
\end{pmatrix}\\ &V_3=\begin{pmatrix}
\frac{a}{a_0} & 0 & 0\\
0& \frac{a}{a_0}&0\\
0&0&\frac{c}{a_0}
\end{pmatrix}\quad.
\end{align}
For clarity, only $V_1$ and $V_2$ are used here in order to describe the entire geometry in the $x$-$y$-plane, but the results are equally valid for the combination of $V_1$ and $V_3$ or $V_2$ and $V_3$.
The lattice vectors of the first variant are given by 

\begin{equation}
\vec{m}_{1,1}=V_1\cdot \vec{a}_1\qquad\vec{m}_{1,2}=V_1\cdot \vec{a}_2
\end{equation}
and for the second variant by
\begin{equation}
\vec{m}_{2,1}=V_2\cdot \vec{a}_1\qquad \vec{m}_{2,2}=V_2\cdot \vec{a}_2\quad.
\end{equation}
To form a twin relation, variant 2 is rotated relative to variant 1 in order to make the  $[110]$ directions of both variants identical. 
\begin{equation}
\label{eq:tw}
V_1\cdot(\vec{a}_1+\vec{a}_2)=Q\cdot V_2.(\vec{a}_1+\vec{a}_2)
\end{equation}
Equation (\ref{eq:tw}) is fulfilled by the rotation matrix $Q$ 
\begin{equation}
Q=\begin{pmatrix}
\frac{2\ctoa}{1+\ctoa^2}&1-\frac{2}{1+\ctoa^2}&0\\
-\frac{2}{1+\ctoa^2}-1&\frac{2\ctoa}{1+\ctoa^2}&0\\
0&0&1
\end{pmatrix}\quad.
\end{equation}
Because this twin boundary connects the smallest variants in the microstructure, it is called "nanotwin boundary" (Fig. \ref{fig:Fig2}) to distinguish them from twins of second or higher order that are usually also present in cubic-to-tetragonal transformations \cite{Roytburd_93}.
To orient the lattice in the usual way with the twin boundaries parallel to the $x$-axis, a further rotation matrix $R$ is defined that is applied to both twin variants, 
\begin{equation}
R=\begin{pmatrix}
\frac{\ctoa}{\sqrt{1+\ctoa^2}} & \frac{1}{\sqrt{1+\ctoa^2}}  & 0 \\
-\frac{1}{\sqrt{1+\ctoa^2}}  &\frac{\ctoa}{\sqrt{1+\ctoa^2}} & 0\\
0 & 0& 1
\end{pmatrix}\quad.
\end{equation}
Since the rotation $R$ acts on the entire lattice, it has no physical consequences and is used for a convenient visualization only. 
\
\subsection{Conversion into the 2M lattice}
The two martensitic variants are redefined to achieve a description in the monoclinic 2M lattice. Using 

\begin{eqnarray}
W_1&=&R\cdot V_1\\
W_2&=&R\cdot Q \cdot V_1
\end{eqnarray}
two new sets of lattice vectors $\vec{n}_{i,j}$ for the two variants are achieved:

\begin{eqnarray}  
\vec{n}_{1,1}&=&(W_1 \cdot \vec{a}_1+W_1 \cdot \vec{a}_2 )/4\\
\vec{n}_{1,2}&=&-(W_1 \cdot \vec{a}_1-W_1 \cdot \vec{a}_2 )/4
\end{eqnarray}
and
\begin{eqnarray} 
\vec{n}_{2,1}&=&(W_2 \cdot \vec{a}_1+W_2 \cdot \vec{a}_2 )/4\\
\vec{n}_{2,2}&=&-(W_2 \cdot \vec{a}_1-W_2 \cdot \vec{a}_2 )/4\quad.
\end{eqnarray}
\MU{As nanotwins can be introduced within a Heusler unit cell, the variants appear not to interact with the chemical order. To obtain translational symmetry in the unit cell of modulated martensite, the stacking period is doubled for all stacking orders with odd stacking parameters (10M instead of 5M and 14M instead of 7M).}
The vectors $\vec{n}_{i,j}$ are sketched in Fig. \ref{fig:Fig2}, which shows the fundamental nanotwin boundary necessary to build a periodically twinned lattice. 

\section{Unit cell of modulated martensite}
\subsection{2M-based description}
The unit cell of modulated martensite is constructed by assuming a stacking sequence of the 2M cells. For example in 10M, this stacking sequence is $(3\bar{2})_2$ (in Zhdanov notation \cite{Nishiyama1978}), described by the integer stacking parameters $d_1=3$ and $d_2=2$, which are repeated twice to ensure chemical order. Here, $d_1$ is shearing to the right (variant 2), and $d_2$ shearing to the left (variant 1). Lattice vectors and parameters of modulated martensite are indexed with a subscript M in the following.

\begin{equation}
\begin{aligned}
\vx{a}{M}{2M}&=2\cdot \vec{n}_{1,1}=2\cdot \vec{n}_{2,1}\\
&=\begin{pmatrix}
\frac{a_0\sqrt{1+\ctoa^2}}{2\ctoa^{\frac13}}\\0\\0
\end{pmatrix}\\ 
\vx{c}{M}{2M}&=\frac{2(d_1 \cdot \vec{n}_{2,2} + d_2 \cdot \vec{n}_{1,2})}{d_1+d_2}\\
&=\begin{pmatrix}
\frac{a_0\left(\ctoa^2-1\right)\left(d_1-d_2\right)}{2\ctoa^{\frac13}(d_1+d_2)\sqrt{1+\ctoa^2}}\\
\frac{a_0 \ctoa^{\frac23}}{\sqrt{1+\ctoa^2}}\\
0
\end{pmatrix} 
\end{aligned}
\end{equation}

The new lattice vectors are named according to the 2M (monoclinic) description of NM, i.e. $\vx{a}{M}{2M}$ is parallel to the nanotwin boundaries, $\vx{c}{M}{2M}$ is along the modulation direction and $\vx{b}{M}{2M}$  is perpendicular to the plane (Fig. \ref{fig:Fig4}).
The monoclinic angle ${\gamma_{\mathrm{\scriptscriptstyle M}}^{
\mathrm{\scriptscriptstyle 2M}}}$ of the structure is given by the vector angle between $\vx{a}{M}{2M}$ and $\vx{b}{M}{2M}$ and is by definition greater than $90^\circ$.
\begin{equation}
\begin{aligned}
&{\gamma_{\mathrm{\scriptscriptstyle M}}^{
\mathrm{\scriptscriptstyle 2M}}}\\&= \pi- \cos^{-1}{\left[\frac{(\ctoa^2-1) (d_1 - d_2)(d_1 + d_2)^{-1}}{\sqrt{
   4 \ctoa^2 + \frac{(\ctoa^2-1)^2 (d_1 - d_2)^2)}{(d_1 + d_2)^2}}}\right]}
\end{aligned}
\end{equation}
%
%

%
	%

\subsection{$\mathrm{L}2_1$-based description}
\begin{figure}[tbp]
	\centering
		\includegraphics[width=\columnwidth]{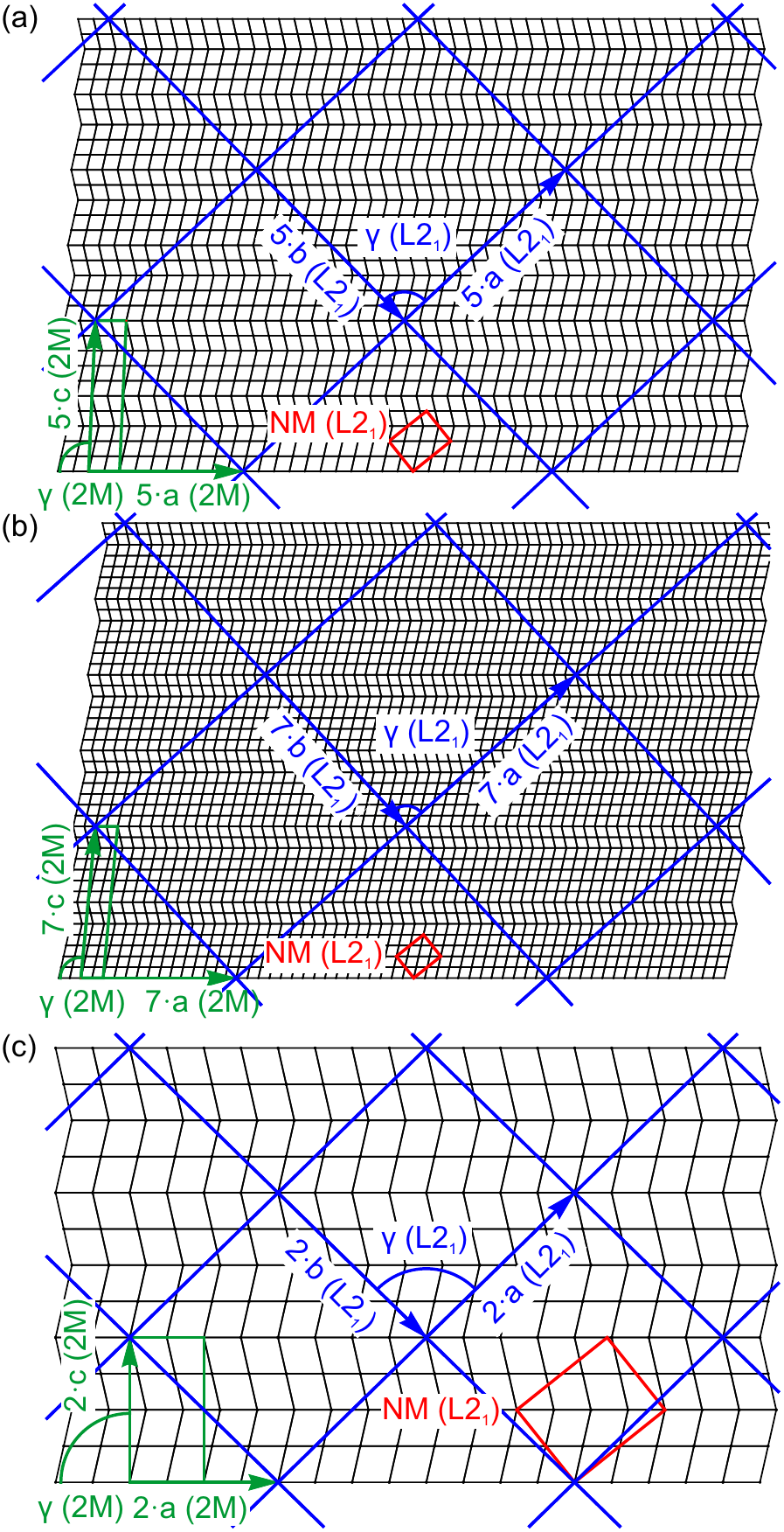}
			
	\caption{\label{fig:Fig3}The (a) 10M (b) 14M and (c) 4O lattices in real space for a high tetragonality of the NM with $\ctoa=1.25$ (red rectangle). The 2M-based monoclinic unit cell (green) of the modulated martensite as well as the larger $\mathrm{L}2_1$ unit cell (blue) are shown including their monoclinic lattice parameters.}
\end{figure}
%
	%
Alternatively, a much larger unit cell can be chosen with a smaller monoclinic angle, which is derived from the cubic $\mathrm{L}2_1$ description of NM,
\begin{eqnarray}
\vx{a}{M}{\mathrm{L2_1}}& =& 
\begin{pmatrix}
\frac{a_0 (\ctoa^2 d_1 + d_2)}{
 \ctoa^\frac13 \sqrt{1 + \ctoa^2} (d_1 + d_2)}\\ \frac{a_0 \ctoa^\frac23}{\sqrt{
  1 + \ctoa^2}}\\ 0\end{pmatrix}\label{eq:L21a}\\
\vx{b}{M}{\mathrm{L2_1}}	&=& \begin{pmatrix}
\frac{a_0 (d_1 + \ctoa^2 d_2)}{
 \ctoa^{\frac13} \sqrt{1 + \ctoa^2} (d_1 + d_2)}\\ -\frac{a_0 \ctoa^\frac23}{\sqrt{
  1 + \ctoa^2}}\\ 0\end{pmatrix}\label{eq:L21b}
\end{eqnarray}

using $t=\ctoa^2$.
%
The lattice vectors of both unit cells can be converted into each other by
\begin{eqnarray}
\vx{a}{M}{\mathrm{L2_1}}&=&\vx{a}{M}{2M}+\vx{c}{M}{2M}\\
\vx{b}{M}{\mathrm{L2_1}}&=&\vx{a}{M}{2M}-\vx{c}{M}{2M}\quad.
\end{eqnarray}
The relation between the cell derived from 2M and the $\mathrm{L2_1}$ cell derived from the cubic lattice is shown in Fig. \ref{fig:Fig3} where the 10M, 14M and 4O lattices are sketched with realistic distances and angles. All lattices follow the same construction principle as 10M, but 14M has a stacking order of $(5\bar{2})_2$ and 4O has a symmetric stacking order of $(2\bar{2})$. 

The $\mathrm{L2_1}$ cell (blue) is related by simple geometry to the monoclinic cell (red). The $\mathrm{L2_1}$ cell also holds translational symmetry. It has an orientation very close to the cubic unit cell of austenite. Additionally, in 10M and 14M, the $b$-axis of the $\mathrm{L2_1}$ cell is almost parallel to the habit plane and to the type I and type II twins, which makes this cell practical to describe structure and microstructure. 
%
%
%
\begin{figure*}[tbp]
	\centering
		\includegraphics[width=\textwidth]{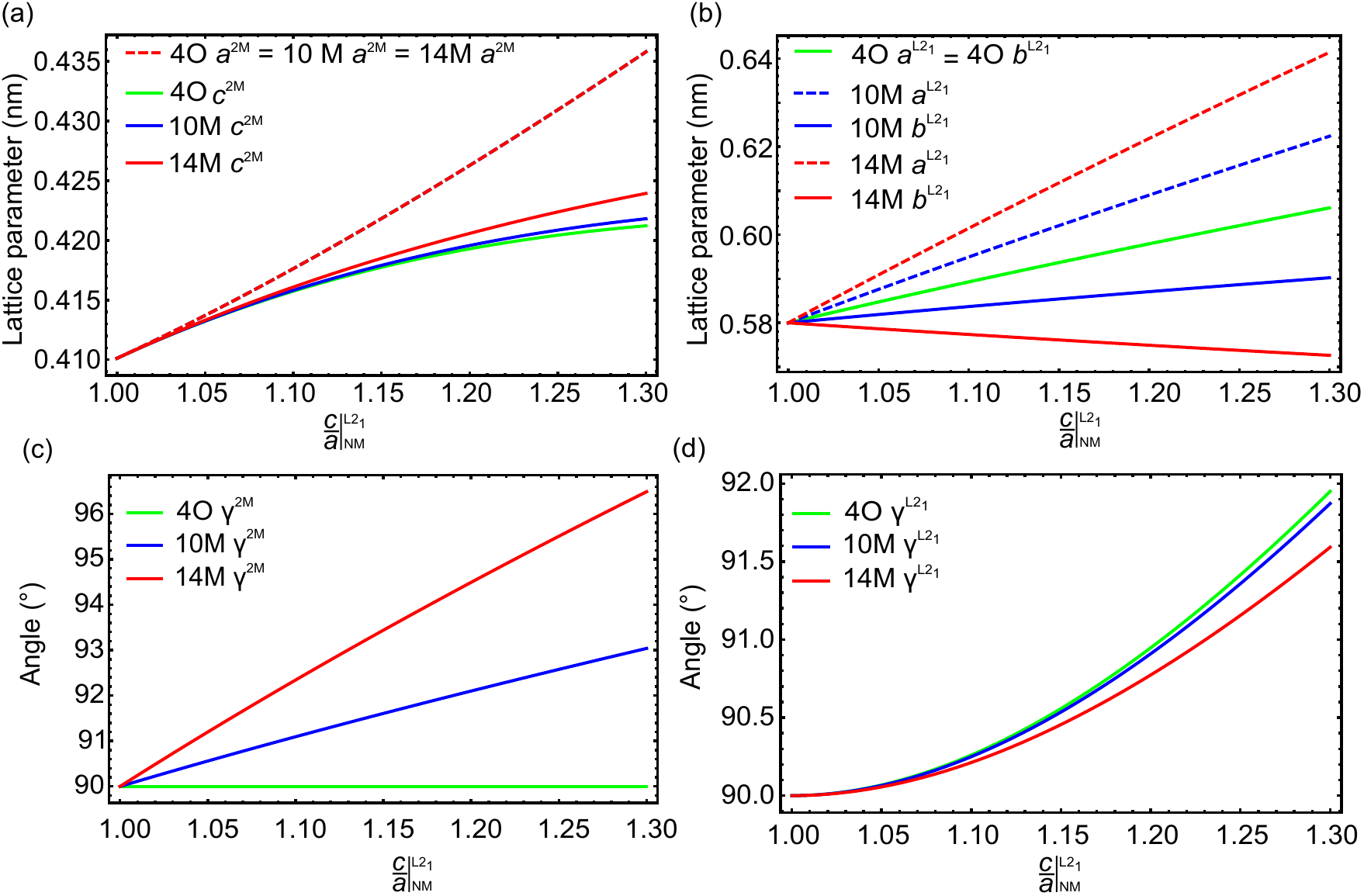}
	\caption{Lattice constants of 10M, 14M, and 4O in dependence of the $\ctoa$-ratio of NM ($\mathrm{L2_1}$). (a) Lattice spacings in $\mathrm{2M}$ description of the modulated phases, (b) in $\mathrm{L2_1}$-description. (c) Monoclinic angle in the $\mathrm{2M}$, and (d) of the $\mathrm{L2_1}$-based lattice.   }
	\label{fig:Fig4}
\end{figure*}
\subsection{Influence of $\ctoa$}
The $\ctoa$-ratio is the fundamental parameter describing the lattice. In the following, it will be shown how the lattice parameters of the resulting cells of modulated martensite depend on $\cta{\mathrm{L2_1}}$.

In Fig. \ref{fig:Fig4}, all lattice constants (in 2M and $\mathrm{L2_1}$ description) are given as a function of $\cta{\mathrm{L2_1}}$ of the fundamental NM lattice. For the calculation, an austenite lattice parameter of $0.58\,\mathrm{nm}$ is assumed, which has no influence on the amount of shearing or tetragonality, only on the absolute values of all lattice parameters.

In the 2M-description (Fig. \ref{fig:Fig4}c), the 4O stacking has a $\gamma$ angle of exactly $90^\circ$, for 10M it is higher and for 14M the highest (up to $96^\circ$), which is a consequence of the higher asymmetry of the stacking sequence.
In contrast to that, in the $\mathrm{L2_1}$ unit cell (Fig. \ref{fig:Fig4}d), 4O has the largest monoclinic angle, while the angles in 10M and 14M are a little smaller. Overall, they do not differ by more than $1^\circ$ and their absolute value is close to $90^\circ$.

$|\vx{a}{4O}{\mathrm{L2_1}}|$ and $|\vx{b}{4O}{\mathrm{L2_1}}|$ are identical for the 4O phase, but $|\vx{a}{4O}{2M}|$ and $|\vx{c}{4O}{2M}|$ slightly differ. Therefore, 4O \MU{is orthorhombic and} $\vx{a}{4O}{2M}$-$\vx{b}{4O}{2M}$-twin boundaries should not be observed \MU{in agreement with the crystallographic theory of martensite.} It will be shown in sec. \ref{ab} how $a$-$b$-twinning reverses the stacking order, which has no effect in 4O. 

For 10M, $|\vx{a}{10M}{\mathrm{L2_1}}|$ is larger than $|\vx{b}{10M}{\mathrm{L2_1}}|$. This difference increases with increasing $\ctoa$. This contradicts the experimental finding that 10M is almost tetragonal.  $|\vx{b}{10M}{\mathrm{L2_1}}|$ is also always larger than $a_0$, which makes it energetically not favorable to form $a$-$b$-twin boundaries at the habit plane in order to reduce stress. \MU{They are often observed in 10M. However, most of these samples are single crystals that were not studied directly after the martensitic transformation, but after additional mechanical treatment, e.g. in order to prepare a type II twin boundary. In this case, a-b-twin boundaries form to adapt to additional stress not originating from the habit plane. In polycrystals, neighbouring grains impose additional  stresses onto each other that also may lead to the formation of a-b-twins. Another driving force may be the ordering of nanotwin boundaries reported recently \cite{Gruner2016}.}

\MU{In contrast to that, in 14M, $a$-$b$-twin boundaries can reduce stress at the habit plane,} where $|\vx{a}{14M}{\mathrm{L2_1}}|$ is much larger than $|\vx{b}{14M}{\mathrm{L2_1}}|$ in agreement with the experimental observations. Also, $|\vx{b}{14M}{\mathrm{L2_1}}|$ is always smaller than $a_0$, while $|\vx{a}{14M}{\mathrm{L2_1}}|$ is larger than $a_0$. $a$-$b$ twin boundaries can therefore form to increase adaption at the habit plane. 
\begin{figure}[tbp]
	\centering
		\includegraphics[width=\columnwidth]{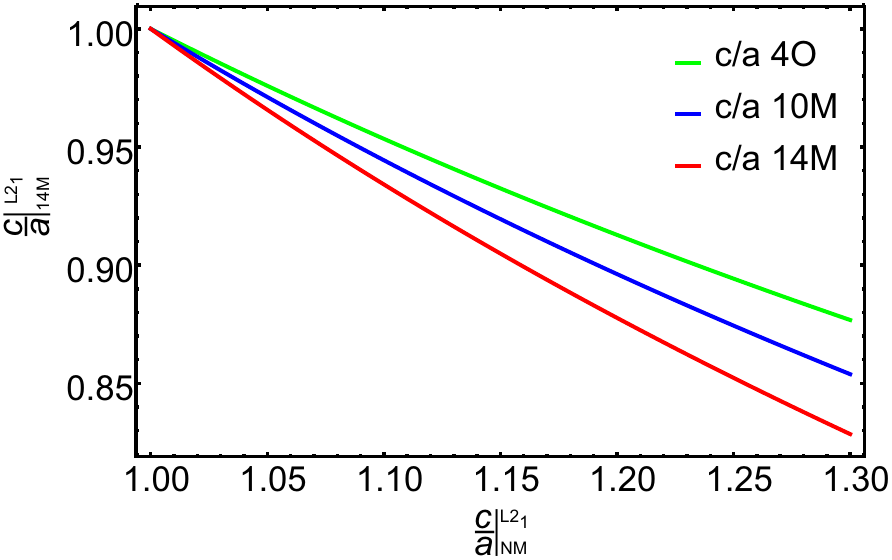}
	\caption{Dependence of $\ctam{L2_1}$ of the modulated structure on $\cta{L2_1}$ of the fundamental NM cell in $\mathrm{L2_1}$ description.}
	\label{fig:Fig5}
\end{figure}

The $\ctam{\mathrm{L2_1}}$-ratio of the modulated phases is plotted in Fig.~\ref{fig:Fig5}. It is always smaller than one and it decreases with increasing $\cta{\mathrm{L2_1}}$-ratio of the fundamental NM cell, and depends on the stacking order. It is the closest to 1 for 4O and the smallest for 14M, which is in agreement with the experimental observations \cite{Pons2000}. 
\section{$a$-$b$-twin boundaries}
\label{ab}
\begin{figure*}[tbp]
	\centering
		\includegraphics[width=\textwidth]{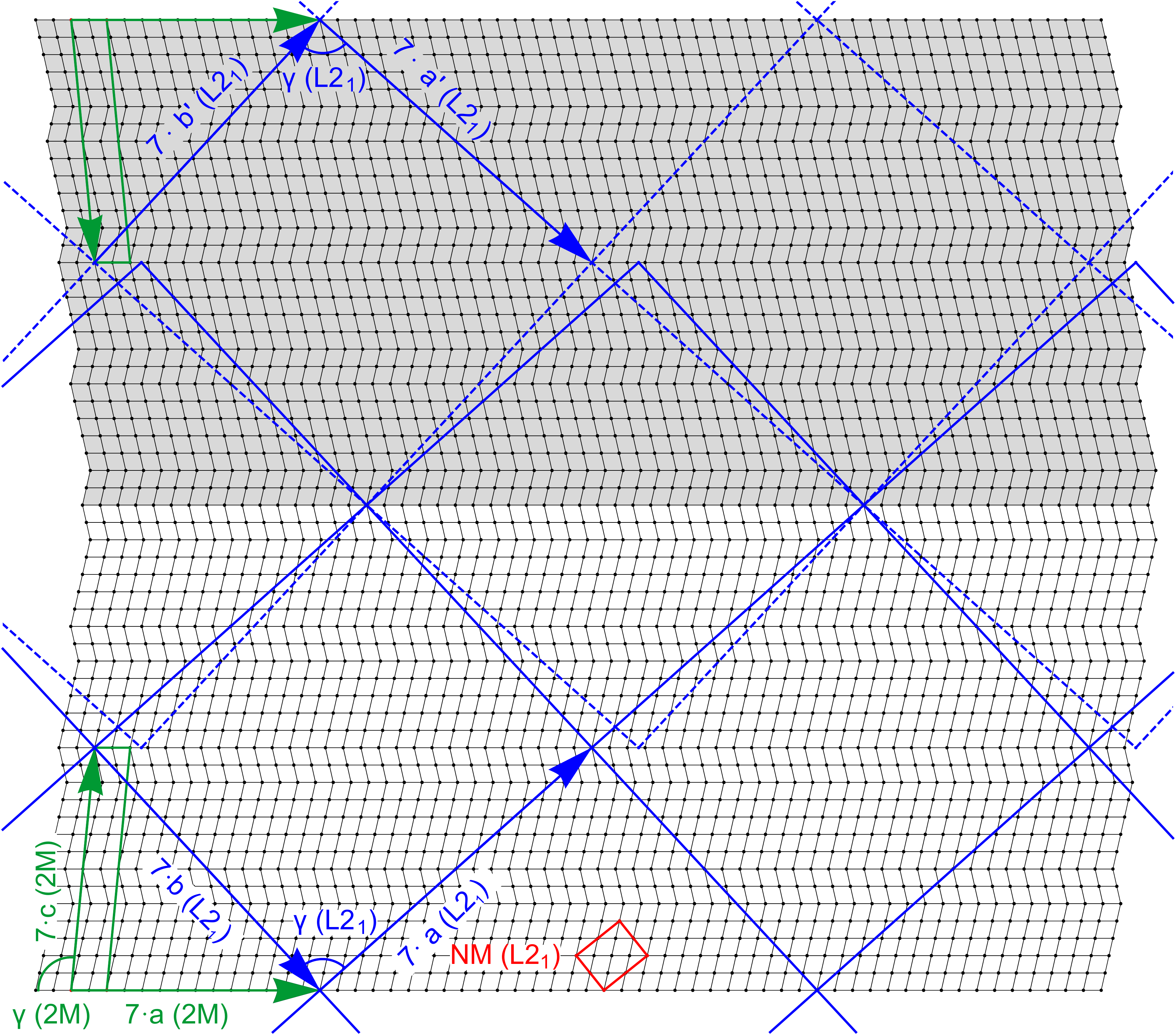}
	\caption{$a$-$b$-twin boundary shown exemplarily for 14M. The top and the bottom half of the lattice have inverted stacking sequences. This leads to a twin relationship between the gray and the white variant, which is obvious when regarding the relation between the $\mathrm{L2_1}$ lattices in the bottom (blue) and top (blue, dashed).}
	\label{fig:Fig6}
\end{figure*}

Multiple types of twin boundaries exist for the modulated structures, including modulation twins, type I and II twins, compound twins and non-conventional twins \cite{Bhat2003,Straka2011b}. Most of them are commonly observed at once. Their presence, orientation and distribution may all influence the critical stress of pseudoplasticity \cite{Heczko2013}. 
Compound  $a$-$b$ twins (always referred to in $\mathrm{L2_1}$-description) are just complex stacking faults because they can equally be described by mirroring the $\mathrm{L2_1}$ cell of martensite or by inverting the stacking sequence:
Fig. \ref{fig:Fig6} shows a 14M lattice in which the stacking order changes from ($5\bar{2}$) to ($2\bar{5}$) in the middle. The $\mathrm{L2_1}$ unit cells for the top and the bottom variants are shown. 
The stacking order is reversed for $\vx{a}{M}{\mathrm{L2_1}}$ and $\vx{b}{M}{\mathrm{L2_1}}$ in equation (\ref{eq:L21a}) and (\ref{eq:L21b}) by exchanging $d_1$ and $d_2$ and renaming $\vx{a}{M}{\mathrm{L2_1}}$ to $\vx{b}{M}{\mathrm{L2_1}}$ and vice versa in order to keep $|\vx{a}{M}{\mathrm{L2_1}}|\geq |\vx{b}{M}{\mathrm{L2_1}}|$.
\begin{eqnarray}
\vx{b'}{M}{\mathrm{L2_1}}&=& \vx{a}{M}{\mathrm{L2_1}}(d_1\leftrightarrow d_2)\\& =& 
\begin{pmatrix}
\frac{a_0 \ctoa^2 d_2 + d_1)}{
 \ctoa^\frac13 (d_1 + d_2) \sqrt{1 + \ctoa^2} }\\ -\frac{(a_0 \ctoa^\frac23}{\sqrt{
  1 + \ctoa^2}}\\ 0\end{pmatrix}\\
\vx{a'}{M}{\mathrm{L2_1}}&=& \vx{b}{M}{\mathrm{L2_1}}(d_1\leftrightarrow d_2)\\	&=& \begin{pmatrix}
\frac{a_0 d_2 + \ctoa^2 d_1)}{
 \ctoa^{\frac13} (d_1 + d_2) \sqrt{1 + \ctoa^2} }\\ \frac{(a_0 \ctoa^\frac23}{\sqrt{
  1 + \ctoa^2}}\\ 0\end{pmatrix}
\end{eqnarray}
This is identical to mirroring $\vx{a}{M}{\mathrm{L2_1}}$ and $\vx{b}{M}{\mathrm{L2_1}}$ at the nanotwin boundary, which is parallel to the $x$-axis, using the reflection matrix
\begin{equation}
M=\begin{pmatrix}
1&0&0\\
0&-1&0\\
0&0&-1\\
\end{pmatrix}\qquad.
\end{equation}
Therefore, 
\begin{eqnarray}
M\cdot\vx{a}{M}{\mathrm{L2_1}}&=&\vx{a'}{M}{\mathrm{L2_1}}\\
M\cdot\vx{b}{M}{\mathrm{L2_1}}&=&\vx{b'}{M}{\mathrm{L2_1}}
\end{eqnarray}
which means that the top and the bottom variant are twin-related. This is only valid for the \MU{compound} $a$-$b$-twin boundaries. In non-conventional $a$-$b$ twins also the modulation direction changes at the twin boundary, hence they are probably not coherent and their description will be far more complex than a simple change of stacking order.  
\section{Austenite-martensite interface}
The martensitic transformation relies on the formation of phase boundaries. Hence, the geometry of the austenite-martensite interface plays an important role for all functional properties arising from the phase transition. The following is based on the Wechsler-Lieberman-Read (WLR) theory \cite{Wechsler1953} that was also developed in parallel by Bowles and Mackenzie \cite{Bowles1954}. A detailed description was published by Bhachattarya \cite{Bhat2003} and Zanzotto and Pitteri \cite{Pitteri1998}.
\subsection{Compatibility criterion}
\begin{figure}[tbp]
	\centering
		\includegraphics[width=\columnwidth]{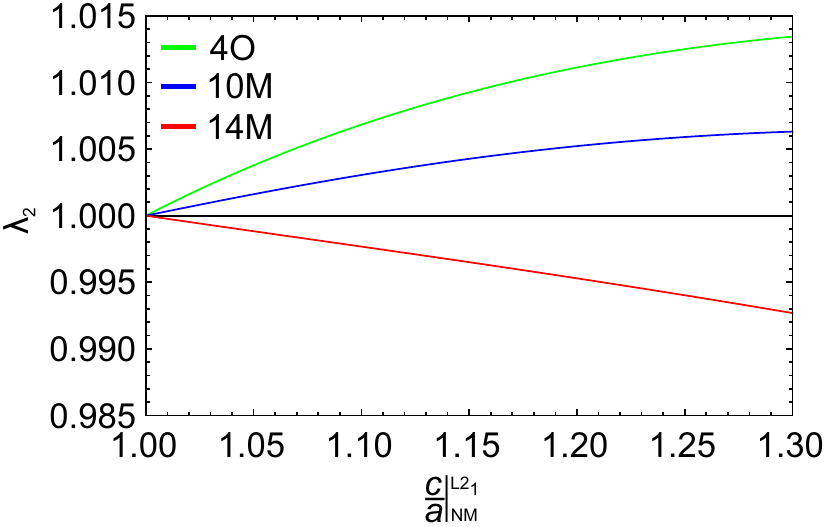}
	\caption{Compatibility of the different adaptive phases in dependence of $\ctoa$ expressed by the middle eigenvalue of the transformation matrix, $\lambda_2$. A value of 1 would be perfect adaptivity.}
	\label{fig:Fig7}
\end{figure}

For a reversible phase transformation, the \MU{elastic} compatibility between the phases is crucial. It can be evaluated by calculating the middle eigenvalue of the transformation matrix in symmetric form  \cite{Ball1987}. If the phase interface is not stress-free but strained, $\lambda_2$ deviates from 1. If the habit plane condition is fulfilled exactly then $\lambda_2=1$, which is one of the prerequisites to very exotic martensitic microstructures shown recently \cite{Song2013}. 
The transformation matrix is given by
\begin{equation}
U=\omega\cdot V_1 + (1-\omega)\cdot Q\cdot V_2\quad.
\end{equation}
It is a weighted sum of both tetragonal variants. The second is rotated to make both \MU{twin-related}, using $Q$ defined above. The weight $\omega$ depends on the stacking sequence 
\begin{equation}
\omega=\frac{d_1}{d_1+d_2}\qquad.
\end{equation}
Because $U$ is not necessarily symmetric, the eigenvalues of the symmetric matrix 
\begin{equation}
C=U^TU
\end{equation}
are calculated. If $\mu_2$ is the middle eigenvalue of $C$, then $\lambda_2$ is given by 
\begin{equation}
\lambda_2=\sqrt{\mu_2}\qquad,
\end{equation}
which is a long but simple algebraic expression of $\ctoa$, $d_1$ and $d_2$. 
The eigenvalue $\lambda_2$ as a function of $\cta{\mathrm{L2_1}}$ is plotted in Fig. \ref{fig:Fig7} for the 10M ($d_1 =3$, $d_2=2$), 14M ($d_1=5$, $d_2=2$) and 4O ($d_1=d_2=2$) phase.
Obviously, the $\lambda_2$ of the 4O is \MU{furthest} from $1$, which means it is the least compatible. 10M and 14M have better compatibility, but are still not perfect. The actual phase boundary will be more complex than \MU{a simple periodic stacking.} Increasing $\ctoa$ always increases the difference between $\lambda_2$ and 1. Note that if volume change is introduced, this picture changes significantly, but this is beyond the scope of this article.

\subsection{Geometry of the habit plane}
To calculate the relative orientation of austenite and martensite, the habit plane condition 

\begin{equation}
\tilde{Q}\cdot U-I=\vec{s}\otimes\vec{h}
\end{equation}
has to be solved. The martensite (represented by $U$) is rotated relative to the austenite (represented by the identity matrix $I$) using the rotation matrix $\tilde{Q}$. $\vec{s}$ is the shear at the habit plane, while $\vec{h}$ is the habit plane normal. 
The habit plane condition can be solved \cite{Bhat2003}. Only $U$ is necessary to calculate $\tilde{Q}$ and $\vec{h}$. These values are functions of $\ctoa$, $d_1$, and $d_2$ only. If the resulting $\tilde{Q}$ is not a pure rotation matrix, it means that the habit plane is strained ($\lambda_2\neq 1$). The pure rotation of martensite relative to austenite is separated by polar decomposition of $\tilde{Q}$.
\begin{figure}[tbp]
	\centering
		\includegraphics[width=\columnwidth]{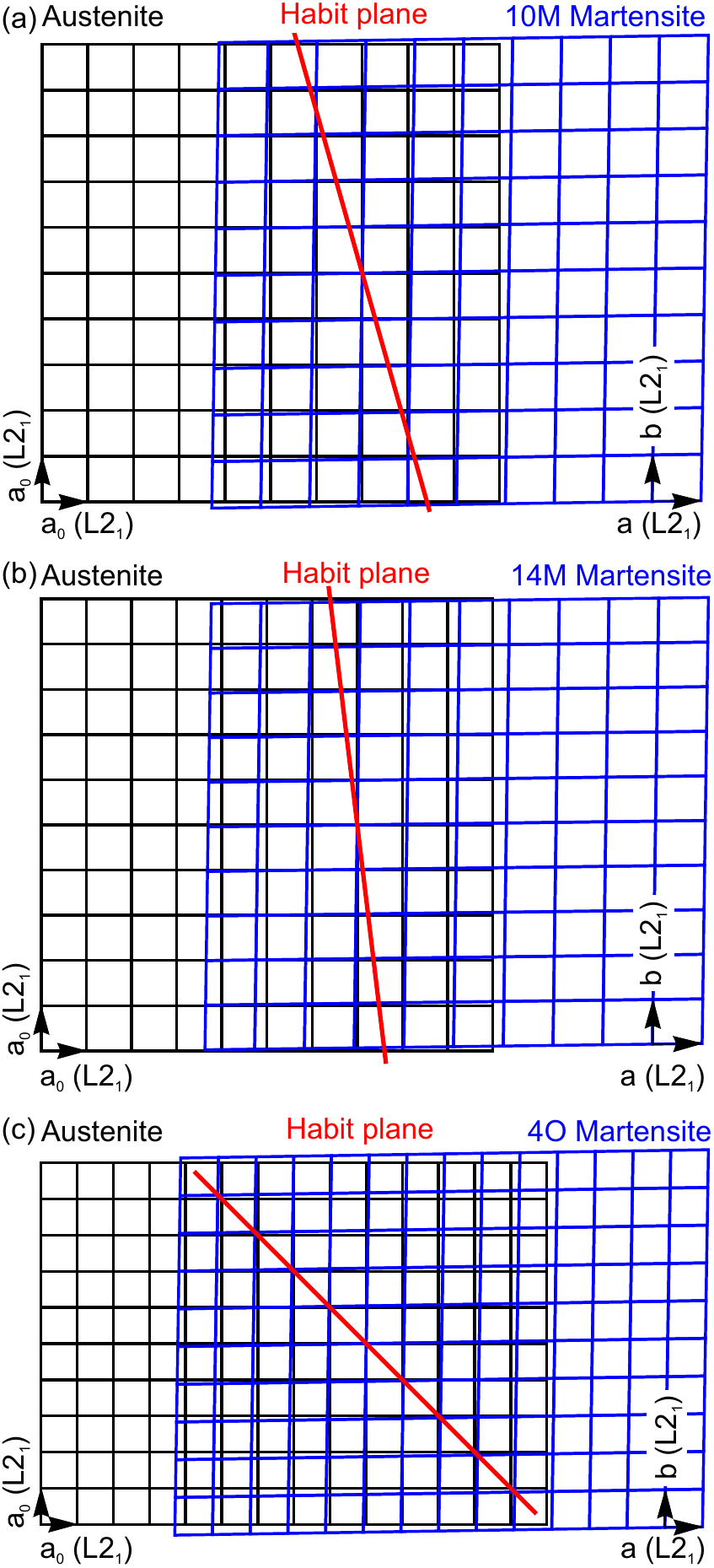}
	\caption{Habit plane between austenite and martensite for (a) 10M, (b) 14M and (c) 4O in real space. The orientations of martensite and of the habit plane were calculated using a high $\ctoa=1.3$ for clarity. The habit plane (red) deviates from any low-indexed lattice plane. Both lattices are shown with a certain overlap to illustrate the misfit.}
	\label{fig:Fig8}
\end{figure}

%
One solution of the austenite-martensite interface is shown in Fig. \ref{fig:Fig8} for (a) 10M, (b) 14M, and (c) 4O. The orientations in the figures were calculated using $\cta{L2_1}=1.3$ to have a clear picture, but there is no qualitative difference when compared to a lower $\ctoa$. Both phases are represented by their unit cells in $\mathrm{L2_1}$-description. The martensite is monoclinically distorted and rotated relative to the austenite. The habit plane orientation is close to $(101)_\mathrm{A}$ (in austenite lattice) which coincides with the line where the least relative distortion between both lattices appears in the figure. The lattices do not match perfectly in the habit plane, which is the consequence of $\lambda_2>1$. While the misorientation between austenite and martensite is less than $1^\circ$ also for this extreme value of $\ctoa$, the habit plane deviates strongly from the $(101)_\mathrm{A}$ plane and the $\vx{b}{M}{L2_1}$-axis, respectively. This leads to the finite aspect ratio of the martensitic nucleus \cite{Bhat2003,Tan1990}.

In 14M, the martensite lattice is contracted relative to the austenite when looking along the habit plane, because $\lambda_2<1$. The deviation of the habit plane from $(101)_\mathrm{A}$ is much smaller than in 10M, but still significant. The absolute misorientation between austenite and martensite is small. The actual stacking sequence directly at the interface \MU{will be more complex, and contain e.g. $a$-$b$-twin boundaries or other stacking faults.} 

The misfit between 4O and austenite is much more pronounced. The habit plane is about $(221)_\mathrm{A}$ for $\ctoa\gtrapprox1$ and close to $(322)_\mathrm{A}$ for $\ctoa\lessapprox1.3$. Recently, it was reported that the 4O forms $\{221\}$ twin boundaries in NiCoMnSn \cite{Lin2016} which is reasonable since habit planes turn into twin boundaries when different nuclei coalesce. This difference to 10M and 14M is a consequence of the symmetric stacking sequence. 4O might not be able to form an interface to \MU{austenite} directly. It is more likely that another stacking order dominates directly at the interface. 


\section{\MU{Conclusion}}
The present article contributes to the ongoing discussion about the nature of the modulated martensites by presenting the exact geometry behind the concept of adaptive martensite for Ni-Mn-X \cite{Kaufmann_PRL10,Khach1991a}. The results were obtained analytically using only the parameter $\ctoa$ of a fundamental NM. From the point of the concept of adaptive martensite, the geometry is valid for all values of the lattice stacking parameters $d_1$ and $d_2$. The rationale of lattice stacking was also used to describe compound $a$-$b$-twin boundaries in 10M and 14M as a simple inversion of the stacking order, and therefore as complex stacking faults. These twin boundaries were often observed in 10M single crystals \cite{Heczko2013} and are a hint that also other, less regular stacking orders should exist \cite{Ustinov2009}. 

In addition, the orientation of the habit plane was calculated using the WLR theory. We found that the habit planes of 14M and 10M are about parallel to $(101)_\mathrm{A}$, but the 4O plane is much closer to a $(221)_\mathrm{A}$ plane. 
All modulated phases are basically not exactly compatible, but 10M and 14M are only half as much strained in comparison to 4O. In 14M, the incompatibility could be resolved by the introduction of $a$-$b$-twin boundaries. In 10M, this is not reasonable because $\vx{a}{10M}{L2_1}$ and $\vx{b}{10M}{L2_1}$ are both longer than $a_0$ and $a$-$b$-twinning can not bring $\lambda_2$ closer towards 1. \MU{It was shown that 4O is always orthorhombic because of the symmetric stacking sequence.} 

%
\section{Acknowledgement}
We thank A. Diestel, B. Schleicher, S. Schwabe, L. Straka, O. Heczko, O. Sozinov, M. Zeleny and R. Chulist for helpful discussions and acknowledge funding by the Deutsche Forschungsgemeinschaft via SPP 1599 www.ferroiccooling.de.
\bibliographystyle{elsarticle-num}
\bibliography{NiMnGa}

\begin{thebibliography}{10}
\bibitem{Supp1}
GIF animations of the 14M, 10M and 4O lattice an a-b-twin bounadry in
  dependence of c/a can be found at [Link].
\end{thebibliography}

\end{document}